\documentclass[a4paper,fleqn]{cas-dc}

\usepackage{subcaption}
\usepackage{float}
\def\tsc#1{\csdef{#1}{\textsc{\lowercase{#1}}\xspace}}
\tsc{WGM}
\tsc{QE}

\begin{document}
\let\WriteBookmarks\relax
\def\floatpagepagefraction{1}
\def\textpagefraction{.001}

\shorttitle{Data-Driven Subgrid Turbulence Modeling}

\shortauthors{Eduardo Vital et al.}

\title [mode = title]{Neural Network for Subgrid Turbulence Modeling for Large Eddy Simulations}

\author[1]{Eduardo Vital}[orcid=0000-0001-6714-6774]

\cormark[1]
\ead{eduardo.vital-brasil-lorenzo-fernandez@ifpen.fr}

\author[1]{Jean-Marc Gratien}[orcid=0009-0005-5143-8082]
\ead{jean-marc.gratien@ifpen.fr}

\author[1]{Yassine Ayoun}[orcid=0009-0008-3448-6180]
\ead{yassine.ayoun@ifpen.fr}

\author[1]{Thibault Faney}[orcid=0009-0005-8879-0012]
\ead{thibault.faney@ifpen.fr}

\author[1]{Julien Bohbot}[orcid=0009-0002-0207-361X]
\ead{julien.bohbot@ifpen.fr}

\affiliation[1]{organization={IFP Energies nouvelles (IFPEN)},
            addressline={1-4 Av. du Bois Préau},
            postcode={92852},
            city={Rueil-Malmaison},
            state={Île de France},
            country={France}}
          
\cortext[1]{Corresponding author}



\begin{abstract}
When simulating multiscale systems, where some fields cannot be fully prescribed despite their effects on the simulation's accuracy, closure models are needed. This phenomenon is observed in turbulent fluid dynamics, where Large Eddy Simulations (LES) depict global behavior while turbulence modeling introduces dissipation correspondent to smaller sub-grid scales. Recently, scientific machine learning techniques have emerged to address this problem by integrating traditional (physics-based) equations with data-driven (machine-learned) models, typically coupling numerical solvers with neural networks. This work presents a comprehensive workflow, encompassing high-fidelity data generation and post-processing, a priori learning, and a posteriori testing, where data-driven models enrich differential equations. 
\end{abstract}


\maketitle

\section{Introduction} \label{sec:intro}

Turbulence is a fundamental aspect of physical phenomena. Its inherently multiscale nature encompasses a vast spectrum of interacting ranges, posing significant challenges for accurate modeling. 

In real-world applications, computational constraints prevent the resolution of small scales as achieved in Direct Numerical Simulation (DNS). Consequently, this unresolved dissipation must be modeled to account for their effects on larger-scale motions, as done in Large Eddy Simulations (LES) through closure models. The closure problem is ubiquitous in systems where macroscopic fields are significantly influenced by finer-scale dynamics. 

In the generic context of multiscale problems described by Partial Differential Equations (PDEs), closure modeling can be systematically formalized as outlined in \cite{BenjaminSanderse2025FoundationsofDataScience}. Consider a typical PDE of the form $F(u, \mu) = 0$, where $F$ is a differential operator, $u(x, t)$ the solution, and $\mu$
 encompasses parameters such as boundary conditions and source terms. Methods aimed at mitigating the computational cost of solving such high-dimensional systems often involve approximating the solution $u(x, t)$ with a lower-dimensional counterpart $\Bar{u}$, achieved through a reduction operator $A: u \mapsto \bar{u}$, representing convolution, projection, or similar techniques. Thus, $\bar{u} \approx A(u)$. Correspondingly, the original system can be simplified as a reduced-order problem $\Bar{F}(u, \mu)$, which serves as an approximation of the true solution $u$ while being computationally more tractable to solve.
 
As the reduction step does not commute with the PDE operator, solving the reduced system demands: $\Bar{F}(u, \mu) = A(F(u, \mu)) = F(\Bar{u}, \mu) + C(u, \Bar{u}, \mu) = 0$, where $C(u, \Bar{u}, \mu)$ represents the closure term. In practice, $C$ cannot be analytically determined from the differential equations, but rather depends on empirical models determined for the application in question.

In our work, we focus on the closure problem of turbulence within the field of Computational Fluid Dynamics (CFD). Here, $u$ represents the velocity field solutions derived from governing equations, such as the Navier-Stokes (NS) or Boltzmann equations, which underpin the high-fidelity approach of Direct Numerical Simulations. In this context, the reduced model employed in Large Eddy Simulations for Navier-Stokes involves solving filtered equations supplemented by a closure component. This addition through a sub-grid scale (SGS) stress tensor $\tau_{ij}$ accounts for the effects of SGS dissipation, representing the turbulent energy dissipated by the unresolved small-scale eddies. Thus, the reduced NS system follows:

\begin{equation} \label{eq:les}
    \frac{\partial \Bar{u}}{\partial t} + \Bar{u} \cdot \nabla \Bar{u} = - \nabla \Bar{p} + \nu \nabla^2  \Bar{u} - \nabla \tau_{ij}
\end{equation}

Where, $p$ is the pressure and $\nu$ the molecular viscosity. Over the years, a variety of models have been developed to approximate $\tau_{ij}$ --- some relying on the Boussinesq hypothesis, which relates it to the strain rate tensor, like Smagorinsky, Dynamic Smagorinsky, and WALE; others following structural or statistical approaches. \cite{pope2001turbulent, inproceedings} provide an overview of both types. 

Here, we propose a data-driven approach where the SGS stress tensor is computed using machine learning techniques. The closure problem is defined as a parameterized problem: $C_{\theta}(u, \Bar{u}, \mu) = \Bar{F}(u, \mu) -F(\Bar{u}, \mu)$,
where the learning process involves determining the optimal parameters $\theta$ that best fit the input-output pairs consisting of fine- and coarse-scale solutions obtained through post-processed DNS data. A neural network is trained on a dataset generated using the Taylor-Green Vortex (TGV) \cite{Brachet_Meiron_Orszag_Nickel_Morf_Frisch_1983} test case (see Section \ref{sec:methodo:simu}). The performance of the model is evaluated in a priori mode, using the generated dataset, an then integrated into an LES numerical solver as the closure model --- the a posteriori mode.

The contribution of the paper is the implementation of an end-to-end workflow that not only addresses practical challenges in dataset generation but also bridges the gap between academic research and industrial practices.

The paper is structured as follows: Section \ref{sec:bibli} reviews related works. Section \ref{sec:methodo} outlines the machine learning methodology we developed and implemented. Section \ref{sec:exp} presents and discusses the results, and finally, Section \ref{sec:conclusion} concludes the paper and suggests directions for future work.

\section{Related works} \label{sec:bibli}

Research for data-driven SGS turbulence modeling has been highly active in recent years, particularly with the integration of neural networks. In \cite{BenjaminSanderse2025FoundationsofDataScience}, the authors provide a review of recent contributions aligned with the closure equation formalism discussed in Section \ref{sec:intro}. 

Many of these contributions adopt a priori learning approaches \cite{dai2023development,yu2022kinetic, Kurz_2022, Kim_Park_Choi_2024}, which involve offline training of models by minimizing closure errors. However, these methods often face stability issues when applied in real-world scenarios, prompting studies such as \cite{Park_Choi_2021, BECK2019108910} to explore alternative strategies. Another avenue of research focuses on a posteriori learning \cite{kim2024generalizabledatadriventurbulenceclosure, List_2022, Beck_2021}, where models are trained by solving reduced equations and minimizing solution errors. Although this approach enhances stability, it incurs higher computational costs.

A third strategy involves hybrid methods, which blend a priori and a posteriori learning to strike a balance between performance and stability. Additionally, reconstruction techniques aim to recover high-dimensional solutions from reduced fields \cite{NISTA2025106498, Fukami_2020, Yuan_2020}. All these methods can be further enhanced by incorporating governing equations either as soft constraints, as demonstrated by Physics-Informed Neural Networks (PINNs) \cite{cuomo2022scientificmachinelearningphysicsinformed, raissi2017physicsinformeddeeplearning}, or as hard constraints, such as in Tensor-Basis Neural Networks (TBNNs), which enforce field invariance through symmetry \cite{LING201622, Wu_2018, BOSE2024107483}.

\section{Methodology} \label{sec:methodo}

Our methodology falls under the a priori learning techniques discussed in Section \ref{sec:bibli}. As we mention in Section \ref{sec:exp}, any possible stability issues occurring in a posteriori testing are yet to be analyzed. According to the turbulence closure problem formalism introduced in Section \ref{sec:intro}, we employ \textbf{OpenFoam} \cite{chen2014openfoam} pimpleFoam solver for both DNS and LES of the incompressible Navier-Stokes equations. The closure problem is modeled as a parameterized problem: $C_{\theta}(u, \Bar{u}, \mu)= \nabla \tau_{ij}$ where $\theta$ are the parameters of the neural network and:

\begin{equation} \label{eq:tau}
   \tau_{ij}(u, \Bar{u}, \mu)
 = \overline{u_i u_j} - \Bar{u_i} \Bar{u_j}
\end{equation}

In this section, we detail the dataset construction process and the reduction operator, outline the learning procedure, and discuss the techniques employed to integrate the trained model into the OpenFoam LES.

\subsection{Test Case and Simulations} \label{sec:methodo:simu}

As a starting point we generate a high-fidelity DNS of the 3d Taylor Green Vortex. This unsteady test case is a classical academic experiment for turbulence analysis. It presents periodic boundary conditions distributed in the faces of a uniformly discretized cube, generating multiple turbulence scales.

We choose to launch experiments at a moderate Reynolds number: 1600. This ensures a complex yet stable turbulent regime, presenting a challenging task for the network. We also ensure a wide range of frequencies, enhancing the likelihood of generalization to lower frequencies in potential downstream testing.

To resolve all turbulence scales and account for dissipation from small eddies, we analyze the Kolmogorov length scale: $\eta = \left( \frac{\nu^3}{\epsilon} \right)^{1/4},$
where $\nu$ is the kinematic viscosity, $\epsilon$ is the energy dissipation rate, and $\eta$ the length scale equivalent to the mesh size. We conclude full DNS can be achieved a uniform discretization of the TGV at 512 resolution ($512^3$ elements).

In addition, we set low-fidelity resolution to 64 (an eighth the elements per axis of the fine-resolution and equivalent to 3 reduction operations). In opposition to a single filter, this highlights the differences between LES and DNS.  As a baseline for comparison with the data-driven model, Smagorinsky turbulence model is used. 

\begin{figure}[pos=t]
    \centering
    \includegraphics[width=0.46\textwidth]{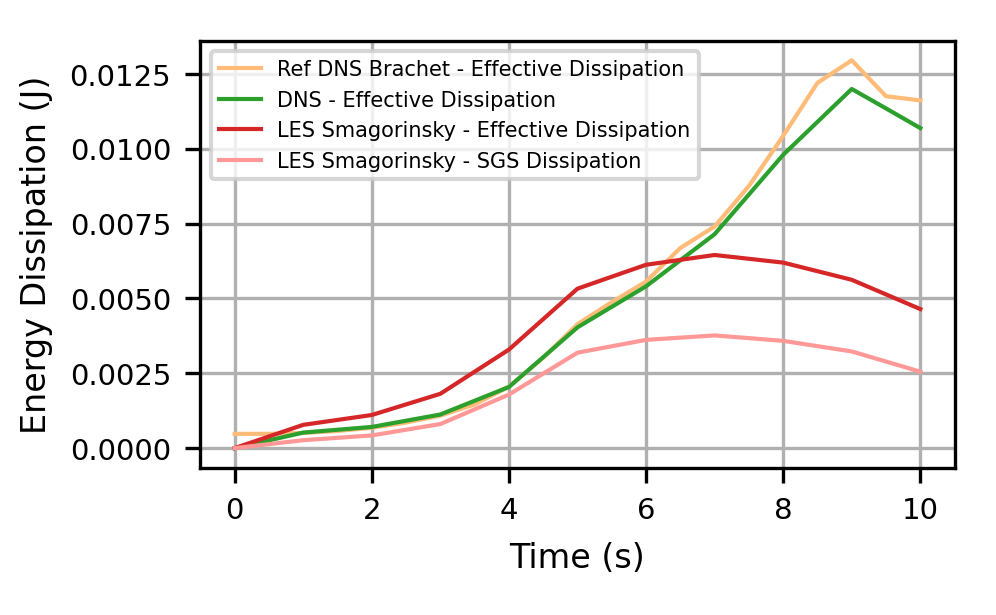}
      \caption{Effective energy dissipation per unit time for the original simulations --- our DNS, the DNS reference from Brachet, and our LES with Smagorinsky turbulence modeling. For the latter, the dissipation due to the turbulence model is also plotted.}
     \label{fig:DeffSimu}
\end{figure}

Both fine and coarse simulations undergo a time convergence study to ensure a maximum Courant number $C = \frac{u \Delta t}{\Delta x}$ of 0.5, where the numerator and denominator represent the temporal and spatial discretizations, respectively. This corresponds to a 0.0025 seconds time step for a converged DNS. Notwithstanding, because of the great volume of data, we write binary fields once per second. We plot on Figure \ref{fig:DeffSimu} the effective dissipation curves during the initial 10 seconds and compare with the literature from Brachet \cite{Brachet_Meiron_Orszag_Nickel_Morf_Frisch_1983}.

\subsection{Processing for Dataset Construction} \label{sec:methodo:dataset}

\begin{figure}[pos=t]
    \centering   \includegraphics[width=0.4\textwidth]{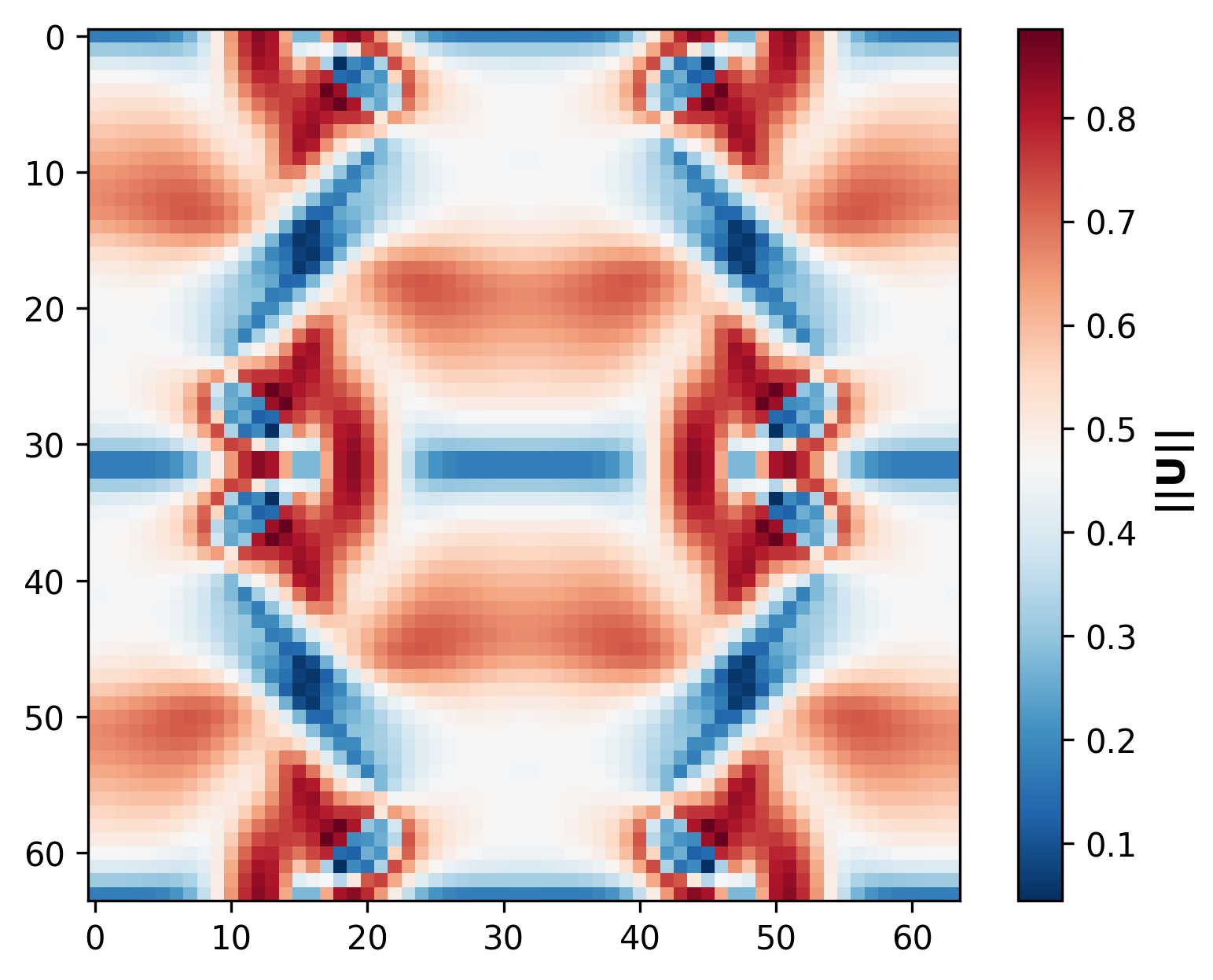}
      \caption{Norm of the filtered field $\Bar{u}$ --- low resolution set at 64. The image represents a cut at the origing according the x-axis at 7 seconds of the TGV simulation.}
     \label{fig:FilteredU}
\end{figure}

Selecting the appropriate reduction operator $A: u \mapsto \bar{u}$ to obtain $\tau_{ij}(u, \Bar{u}, \mu)$ from \ref{eq:tau} is a non-trivial task, as in LES it is usually implicitly computed through the reduced equations \ref{eq:les}. Common choices of LES filters are mean convolution or, as in the Dynamic Smagorinsky implementation \cite{alberto_passalacqua_2021_4697995} for OpenFoam, a cell-averaging approach. For data-driven approaches, however, the reduction operator must allow derivatives coherent with a coarse grid while preserving the momentum; hence, we adopt a block reduction operator similar to the approach employed in \cite{Kim_Park_Choi_2024}. Figure \ref{fig:FilteredU} illustrates the coarse ground-truth solution.

We further extract the symmetric components and apply a deviatoric operation to $\tau_{ij}$, i.e., $\boldsymbol{\tau_{ij}}^d = \boldsymbol{\tau_{ij}} - \frac{1}{3}\text{tr}(\boldsymbol{\tau_{ij}}) \cdot \mathbf{I}$. In this way, we extract only the shear, anisotropic part of the tensor, removing the isotropic part that can be accounted by the pressure term in \ref{eq:les}. A similar approach is done by \cite{alberto_passalacqua_2021_4697995} on the strain rate tensor $S_{ij}$. Gradients, in its turn, are computed with centered finite differences.

\begin{figure}[pos=t]
     \centering
     \includegraphics[width=0.46\textwidth]{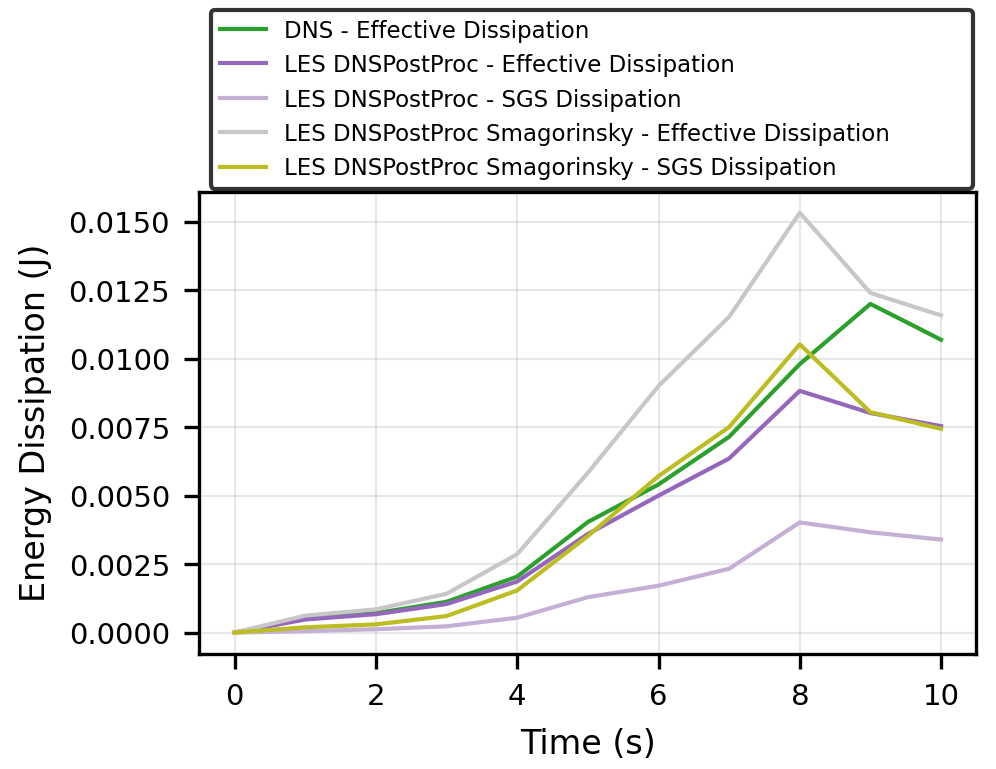}
      \caption{Effective energy dissipation per unit time for original and post-processed simulations --- our DNS and the reduced simulations. The latter were achieved through a reduction operator $A: u \mapsto \bar{u}$ and include different turbulence models (Smagorinsky and \ref{eq:tau}), with their respective dissipation contributions also plotted.}
     \label{fig:DeffPostProc}
\end{figure}

Through the explicit reduction of the velocity, we can not only build the SGS stress tensor but also emulate the behavior of a lower-resolution LES. Figure \ref{fig:DeffPostProc} illustrates the effective energy dissipation of these processed DNS, transformed into LES. We compare equation \ref{eq:tau} to the Smagorinsky model, showing how it over-estimates the energy dissipation.

\subsection{Architecture and Learning Procedure} 
\label{sec:methodo:architecture}

Among the networks discussed in Section \ref{sec:bibli}, we initially selected a simple architecture similar to \cite{Kim_Park_Choi_2024}. It consists of an MLP that performs pointwise inference on the fields, meaning it processes and predicts a single tensor. Therefore, the rich input information of the gradient of the reduced field $\nabla \Bar{u}$ with its 9 components is imputed and the symmetric SGS stress tensor (6 components) is outputted.

To enhance the model's learning capacity, particularly given the scarcity of data due to downsampling to a 64 resolution, we extend the initial shallow network into a deeper architecture by optimizing the number of layers and hidden units during the validation phase. The resulting MLP consists of 9 layers of 1024 neurons. Additionally, a \textit{max absolute scaling} is used to preprocess the data, preserving data sparsity and physical properties in opposition to normalization.

We perform, then, a priori training with hyperparameter optimization to find the best learning scheme, using a classical partition of 60-20-20 for train-test-validation. We employ classical mean squared error (MSE) as loss function, while as evaluation metric, we prefer a relative metric as the R2 score: $R^2_j = 1 - \frac{\sum_{i=1}^N \left( T^{(i)}_j - \hat{T}^{(i)}_j \right)^2}{\sum_{i=1}^N \left( T^{(i)}_j - \bar{T}_j \right)^2}$, where $\mathbf{T}^{(i)}$ and $\hat{\mathbf{T}}^{(i)}$ are, respectively, the predictions and the ground truth for the target tensor of the $i_{th}$ point. We add the mean absolute error (MAE) to complement this metric: $\mathrm{MAE}_j = \frac{1}{N} \sum_{i=1}^N \left| T^{(i)}_j - \hat{T}^{(i)}_j \right|$. The components $j$ of the tensor can also be averaged to achieve a global metric.

\subsection{A Posteriori CFD-AI coupling}
\label{sec:methodo:coupling}

Finally, the model trained a priori is linked to an LES. Trained in simple precision, it is converted to double precision, serialized and compiled with Pytorch's TorchScript, to be loaded by TorchLib in C++ and imported in a custom-compiled version of the CFD software. Additionally, a dedicated GPU inference library is developed, allowing efficient model querying even within parallel OpenFoam environment using MPI. CUDA's scheduler ensures that parallel requests from different processes, each with its own CUDA context, are serialized to prevent memory concurrency.


Scaling and descaling are performed directly within OpenFoam rather than by the model. Scaling parameters are imported by the master rank and broadcast. The transformation is done efficiently when communicating with the inference module --- which requires the bufferization of inputs and field construction from the output buffer. 

We also implement an explicit SGS stress equation, allowing to directly retrieve the model predictions. Specifically, we decouple turbulent viscosity ($\nu_t$) and molecular viscosity ($\nu$) in the calculation. Typically, $\nu_t$ is updated / corrected each iteration and retrieved in the equation:
$D_{eff} = -\nabla \cdot \left(\nu_{\text{eff}} \, \text{dev}({\nabla \Bar{u}}^T)\right) - \nabla^2(\nu_{\text{eff}} \Bar{u})$, where $\nu_{\text{eff}}=\nu + \nu_{t}$, $dev$ represents the deviatoric operation, and $D_{eff}$ denotes the effective diffusion term, encompassing both turbulent and viscous contributions. Thus, we employ: $D_{eff} = -\nabla \cdot \left(\nu \, \text{dev}({\nabla \Bar{u}}^T)\right) - \nabla^2(\nu \Bar{u}) - \nabla\tau_{ij}$. The effective diffusion term is retrieved in the solver's momentum equations, replacing the corresponding terms of PDE \ref{eq:les}.

\section{Experiments} \label{sec:exp}

In this section, we evaluate the model's performance throughout the simulation, leveraging the dataset generated (Sections \ref{sec:methodo:simu} and \ref{sec:methodo:dataset}) and applying the learning scheme and evaluation metrics described in Section \ref{sec:methodo:architecture}.

\begin{figure}[htbp]
    \centering
    \includegraphics[width=0.46\textwidth]{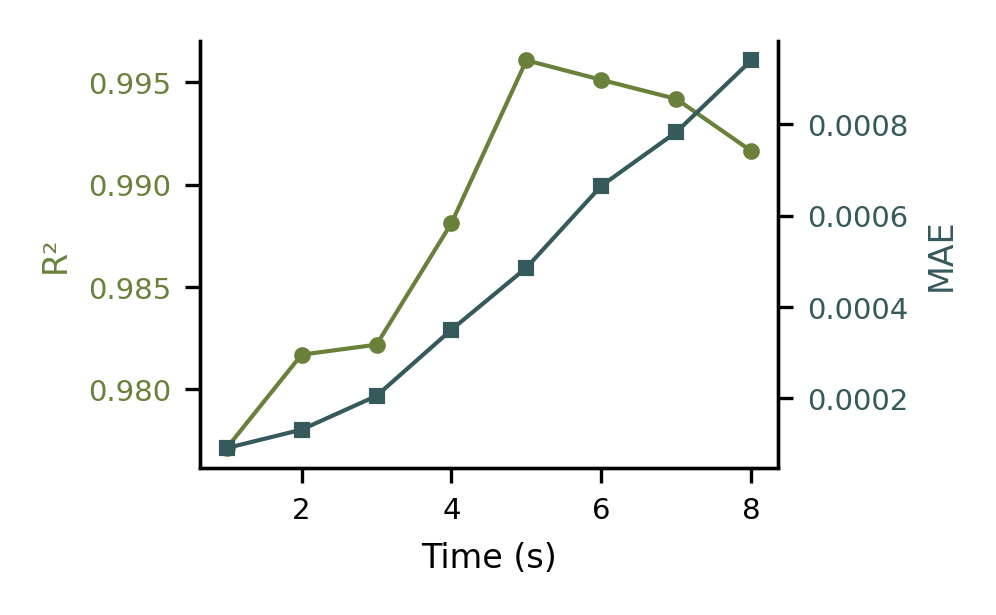} 
    \caption{Evolution of error metrics (MAE and R2 score) over time in a priori, testing setting for the 10 seconds of the TGV experiment. The data originates from a DNS, processed through a reduction operator $A: u \mapsto \bar{u}$.}
    \label{fig:Error Evolution}
\end{figure}

Figure \ref{fig:Error Evolution} illustrates the a priori phase where the model, trained on the entire trajectory, performs on testing in the different time steps. The complementary metrics (MAE and R2 score) demonstrate the excellent performance both on the laminar regime (before 4 seconds), and on turbulent regime, 4 seconds onward. Due to increased dissipation in the later stage (see Figure~\ref{fig:DeffSimu}), absolute errors tend to grow, whereas relative errors are smaller, resulting in a higher R2 score. Conversely, during the initial stage, the R2 value decreases because of the smaller magnitude of the predicted variables; however, this degradation is negligible, as evidenced by the low absolute errors.

We refrain from including the a posteriori analysis in this brief article. Despite promising, it requires further investigation.

\section{Conclusions and Future Work} \label{sec:conclusion}
In this work, we present an end-to-end application for developing a data-driven closure model for turbulence in Large Eddy Simulations. We address practical questions arising during dataset generation and combine academic research with industrial practices to enrich the workflow.

The shortcomings of the present research will be investigated in future work. We recognize the limitations stemming from using a pointwise model. Thus, we intend to implement and test multi-hierarchical approaches such as GNNs and CNNs.
Moreover, the inherent generalisation deficiency of standard neural networks signalizes the importance of more robust approaches --- neural operators for space and time discretization invariance, as well as physically informed techniques for extrapolation to unknown frequencies and test cases. A robust set of experiments and ablation studies, guided by a posteriori performance, enlightens a promising direction.

\bibliographystyle{elsarticle-num}
{\small\bibliography{ML4turbBib}}

\end{document}